\documentclass[10pt,twocolumn,letterpaper]{article}

\usepackage{wacv}              

\usepackage{graphicx}
\usepackage{amsmath}
\usepackage{amssymb}
\usepackage{booktabs}
\usepackage{float}
\usepackage{stfloats}
\usepackage[accsupp]{axessibility} 

\usepackage[pagebackref,breaklinks,colorlinks]{hyperref}
\usepackage{afterpage}
\usepackage{marvosym}

\usepackage[capitalize]{cleveref}
\crefname{section}{Sec.}{Secs.}
\Crefname{section}{Section}{Sections}
\Crefname{table}{Table}{Tables}
\crefname{table}{Tab.}{Tabs.}

\def\wacvPaperID{1940} 
\def\confName{WACV}
\def\confYear{2025}

\begin{document}
\title{MS-Glance: Bio-Inspired Non-semantic Context Vectors and their Applications in Supervising Image Reconstruction}
\author{%
\textbf{Ziqi Gao}$^{1,2}$ \qquad \textbf{Wendi Yang}$^{1,2}$\qquad \textbf{Yujia Li}$^{3}$ \qquad \textbf{Lei Xing}$^{4}$ \qquad  \textbf{S. Kevin Zhou}$^{1,2,3,5}$\thanks{Corresponding author.}   
\\
 $^1$School of Biomedical Engineering, Division of Life Sciences and Medicine, 
 ~~ \\
 University of Science and Technology of China (USTC), Hefei Anhui, 230026, China
  ~~ \\ 
  $^2$Center for Medical Imaging, Robotics, Analytic Computing \& Learning (MIRACLE), 
  ~~ \\
  Suzhou Institute for Advance Research, USTC, Suzhou Jiangsu, 215123, China
 ~~ \\ 
  $^3$Key Laboratory of Intelligent Information Processing of Chinese Academy of 
  ~~ \\
  Sciences (CAS), Institute of Computing Technology, China
 ~~ \\ 
   $^4$ Department of Radiation Oncology, Stanford University, Stanford, CA, USA
 ~~ \\ 
  $^5$Key Laboratory of Precision and Intelligent Chemistry, USTC, Hefei Anhui, 230026, CAS
  ~~ \\ 
 \tt\small \{gaoziqi,yangwendi\}@mail.ustc.edu.cn  yujia.li@miracle.ict.ac.cn \\
 \tt\small lei@stanford.edu s.kevin.zhou@gmail.com
 \vspace{-4pt}
 }

\maketitle

\begin{abstract}
\vspace{-4pt}
Non-semantic context information is crucial for visual recognition, as the human visual perception system first uses global statistics to process scenes rapidly before identifying specific objects. However, while semantic information is increasingly incorporated into computer vision tasks such as image reconstruction, non-semantic information, such as global spatial structures, is often overlooked. To bridge the gap, we propose a biologically informed non-semantic context descriptor, \textbf{MS-Glance}, along with the Glance Index Measure for comparing two images. A Global Glance vector is formulated by randomly retrieving pixels based on a perception-driven rule from an image to form a vector representing non-semantic global context, while a local Glance vector is a flattened local image window, mimicking a zoom-in observation. The Glance Index is defined as the inner product of two standardized sets of Glance vectors. We evaluate the effectiveness of incorporating Glance supervision in two reconstruction tasks: image fitting with implicit neural representation (INR) and undersampled MRI reconstruction. Extensive experimental results show that MS-Glance outperforms existing image restoration losses across both natural and medical images. The code is available at \url{https://github.com/Z7Gao/MSGlance}.
\vspace{-15pt}
\end{abstract}

\begin{figure}
    \centering
    \includegraphics[width=0.95\linewidth]{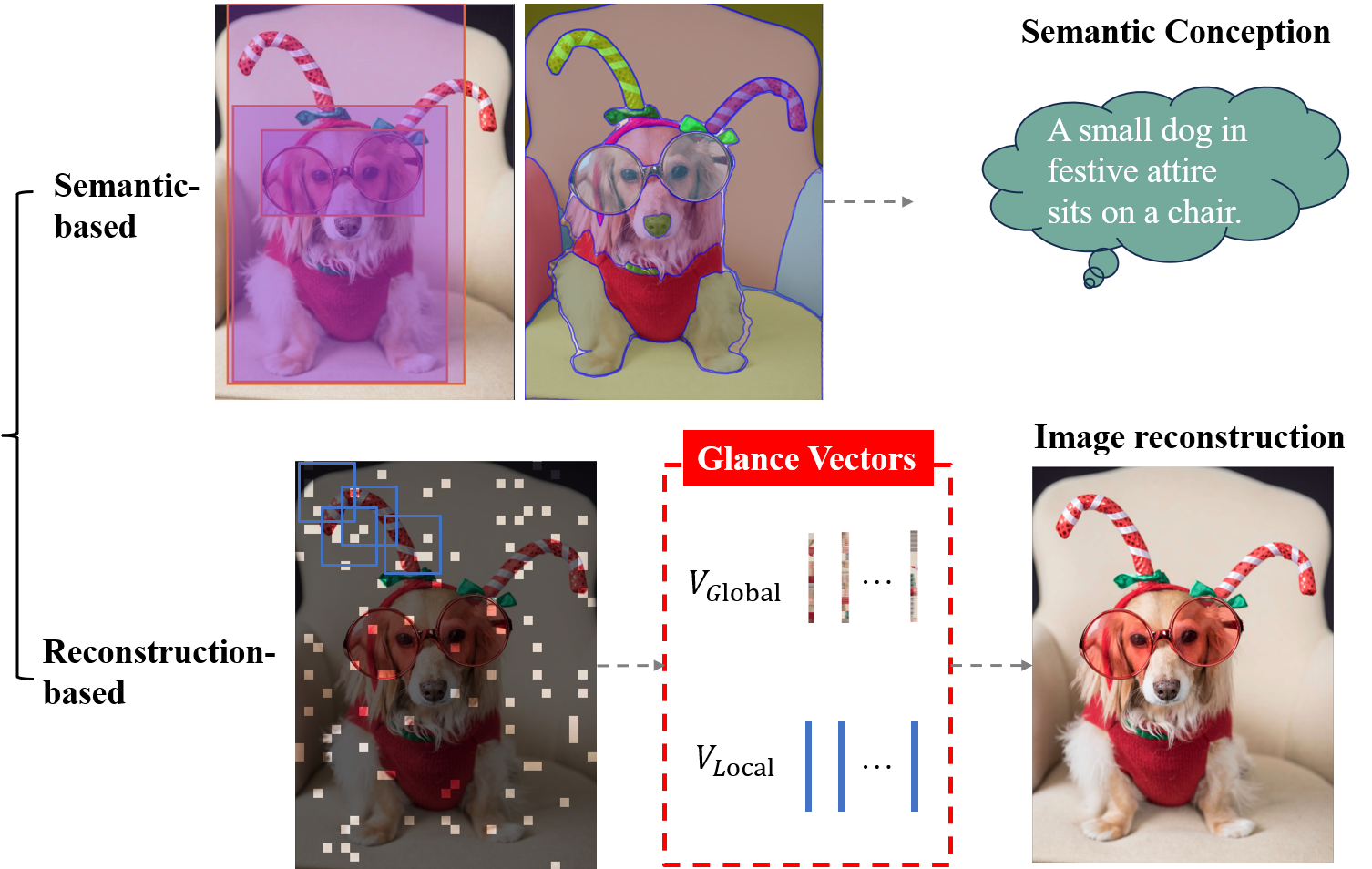}
    \caption{Category of human recognition\cite{Oliva2001ModelingSpatialEnvelope}
    and the extraction of multi-scale Glance Vectors.}
    \label{fig:1}
    \vspace{-15pt}
\end{figure}
\section{Introduction}
Computer vision (CV) has increasingly incorporated global semantic information, inspired by the human visual perception process~\cite{visual_perception} where understanding is driven by high-level semantic cues rather than focusing on individual pixels.
For example, the Vision Transformer (ViT)~\cite{dosovitskiy2021ViT}, a prominent neural network architecture, captures long-range dependencies within an image and generates powerful feature representations through vision-language pretraining\cite{CLIP}. 
In image reconstruction, advanced models have also successfully incorporated semantic information. For instance, in super-resolution, architectures such as U-Net utilize multiple layers of convolutional and downsampling operations to progressively extract high-level semantic information~\cite{unet, unet_sr, unet_sr2}. Besides, in undersampled MRI reconstruction, some researchers utilize semantic segmentation networks to assist in the reconstruction process~\cite{FR_Net, segmentation_aware_mri, segmentation_aware_mri2}.
\label{sec:intro}
However, in human vision, semantic conception\cite{marr_vision} is not the only way to portray the recognition; it can also be interpreted as a reconstruction procedure of the scene that doesn't involve semantic understanding\cite{Oliva2001ModelingSpatialEnvelope,potter1975meaning,Schyns_Oliva_1994,oliva2007role}, demonstrated in Fig.\ref{fig:1}. A pioneering study\cite{potter1975meaning} shows that human vision gleans a large amount of meaningful information from one single glance. Experimental studies followed\cite{Schyns_Oliva_1994,oliva2007role}, suggesting that recognition of the real world may start from encoding global configuration since no perception of individual objects or detailed features can be made in such a short time. A representative computational model, \textit{Spatial Envolope}\cite{Oliva2001ModelingSpatialEnvelope}, models the structure of real-world scenes by a set of perceptual properties including naturalness, openness, roughness, ruggedness, and expansion. Further, Michelle Greene et al. \cite{glance2009} experimentally shows that statistical and structural cues extracted from very brief (e.g. 19 ms and masked) exposures allow for above-chance categorization of scenes, verifying the \textit{Spatial Envelope} model and the existence of a reconstruction-based recognition.

Current image reconstruction algorithms primarily focus on pixel-wise similarity or high-level semantic information, often overlooking non-semantic, statistical, and structural information. Although Structural Similarity Index Measure (SSIM) ~\cite{ssim} is widely used for evaluating image quality, it only captures local information from neighboring pixels without accounting for structural information from distant pixels. 
On the other hand, S3IM~\cite{s3im} improves SSIM in the application of novel view synthesis with Neural Radiance Field (NeRF)~\cite{Nerf} by applying SSIM to random patches. However, the Gaussian-kernel-based computation of SSIM assumes weighting the central pixels while suppressing the edge pixels, limiting its ability to represent global context, as global context does not inherently focus on the center. 

To bridge this gap, we propose Multi-Scale Glance (\textbf{MS-Glance}), a novel non-semantic descriptor of image context, inspired by the human recognition process that bypasses the semantic concept. MS-Glance includes local and global Glance vectors. A global Glance vector is formulated by randomly retrieving pixels from an image with an explicit rule and a local Glance vector is formed by flattening a local window of an image. Given two Glance vectors, their Glance Index is the inner product between two sets of standardized Glance vectors. Since \textbf{MS-Glance} is a descriptor of image context, it can be seamlessly integrated into existing image reconstruction models as a plug-and-play component, enhancing the quality of the reconstructed images. 

We show the applicability of MS-Glance loss in two scenarios: image fitting with implicit neural representation (INR) and supervised undersampled MRI reconstruction with DRDN\cite{frequency_loss2}. For image fitting, we training SIREN~\cite{sitzmann2019siren}, a neural representation capable of modeling signals with fine details, with MS-Glance loss. Using MS-Glance leads to the best SIREN representation ability on datasets of common objects, human faces, and MRI brain scans. For undersampled MRI reconstruction, we propose a novel air prior that allows rule-based pixel selection for MS-Glance. Experiments are conducted on two public datasets, IXI and FastMRI, encompassing various MRI acquisition scenarios and different organs. Extensive experimental results demonstrate that incorporating Glance not only enhances the performance of existing models but also improves image reconstruction quality for both natural and medical images.

The contributions of this work are as follows. First, we propose a biologically inspired non-semantic context descriptor, MS-Glance, along with the Glance Index Measure and Glance loss specifically designed for image comparison. Second, we demonstrate its ability to improve learned image representation through INR image fitting. Third, we apply it to training undersampled MRI reconstruction networks, showcasing its utility in image restoration tasks. Additionally, we introduce a novel perception prior, MRI air prior, and incorporate it with the construction of MS-Glance vectors for MRI reconstruction. Finally, extensive experiments on a wide range of datasets show that the MS-Glance loss outperforms existing loss functions used in image restoration, such as L1+SSIM\cite{image_restoration_review} and LPIPS\cite{MRI-Loss-Hupeng}.

\section{Related work}
\subsection{Image non-semantic information}
While semantic information has been extensively utilized in various CV tasks, research on non-semantic information of images, such as structural and layout features, remains limited. Cao et al.~\cite{non_semantic_facial} introduced the concept of Non-Semantic Facial Parts (NSFP), which identifies the most discriminative patches for face recognition and retrieval. However, their method relies on predefined features like SIFT, limiting the performance. Murrugarra et al.~\cite{non_semantic_transfer} proposed non-semantic transfer from attributes that may belong to different domains, but focuses solely on texture, which is only a small subset of non-semantic information. In autonomous driving, Anas et al.~\cite{localisation_vehicles_non_semantic} represented point clouds with non-semantic features for environment interpretation, localization, and mapping; yet their work is specific to point clouds rather than images. In NeRF, Xie et al.~\cite{s3im} applied SSIM on reorganized random image patches. Still, their work is also application-specific and relies on the original pixel-based SSIM computation with a local focus center.

\subsection{Image reconstruction}
Image reconstruction~\cite{image_sr_review, mri_recon_review, image_restoration_review} in CV refers to recovering or restoring an image from incomplete, corrupted, or undersampled data. Here, we introduce one reconstruction method and one reconstruction task: Implicit Neural Representation (INR) and undersampled MRI reconstruction. 
\subsubsection{Implicit Neural Representation (INR)} INR is a neural network-based continuous image representation. It parameterizes a field in a coordinate-based manner. 
Various research on INR explored better image fitting methods, for example, Sinusoidal Representation Networks (SIREN)~\cite{sitzmann2019siren} utilizes periodic activation functions for INR to capture intricate natural signals and their derivatives. INR has also been used for image compression~\cite{INR_image_compression, INR_image_compression1,Girish_2023_ICCV}, segmentation~\cite{INR_seg}, and super-resoultion~\cite{INR_sr, INR_sr2}. Other major application of implicit neural representation is a series of works~\cite{Nerf, wang2021neus, zhang2020nerf++,gao2024planar} on 3D reconstruction. Instead of using explicit representation(\textit{e.g.}, voxels), NeRf~\cite{Nerf} encodes the representation of 3D object/scene into MLPs. In this work, we focus on improving INR's image-fitting performance, potentially applicable to various related tasks.

\subsubsection{Undersampled MRI reconstruction}
MRI is a non-invasive in-vivo imaging modality that benefits radiological diagnosis. However, its long acquisition leads to patient discomfort and motion artifacts. Undersampling in the K-space domain is often employed to accelerate the acquisition, but this leads to a loss of image fidelity. Parallel Imaging (PI) \cite{PIgriswold2002GRAPPA,PIpruessmann1999SENSE} and Compressed Sensing (CS)\cite{CSlustig2008compressed,CShaldar2011random,CSliang2009accelerating,CSpatel2011gradient} is limited to an acceleration rate of 2-3\cite{CSlustig2008compressed}. Supervised deep learning-based undersampled MRI reconstruction networks are developed to learn the mapping between undersampled images and high-quality MR images and achieve improved performance\cite{eo2018kikiDL,qin2018dc1DL,ding2022mriDL,putzky2019rimDL,wang2016DL-unet,yang2017daganDL,jin2017unetmriDL,schlemper2017dcDL,ding2019DL-rdnunet,huang2022DL-swinmr,sriram2020DL-e2evar,zhang2019reducingDL,quan2018DL-compressedgan,cheng2019spdDL}. Compared with diffusion and GAN-based unsupervised models~\cite{diffusion_recon1,diffusion_recon2,huang2023cdiffmr,chung2022ccdf-diff,nips2021score-diff,Cao2024HFS-SDE-diff,Levac2023motion-diff,ozturkler2023red-diff,chung2022scoreMRI-diff,ozturkler2023smrd-diff,GUNGOR2023AdaDiff,peng2022DiffuseReco,gao2024mrpdundersampledmrireconstruction,GAN_recon, GAN_recon1}, supervised models typically show strong performance in in-domain scenarios. Most works have adopted the commonly used L1, L2, adversarial loss, SSIM loss, or frequency domain loss~\cite{frequency_loss, frequency_loss2, frequency_loss3}. A small amount of work compares the effect of loss functions used for training\cite{MRI-Loss-Hupeng} and explores new loss function\cite{frequency_loss3} that is more effective than existing loss for phase reconstruction. In our work, we focus on the reconstruction of image magnitude, the final part that is displayed in diagnostics, and compare ours with those losses that are proved superior in magnitude reconstruction: SSIM\cite{frequency_loss3} and perceptual loss \cite{MRI-Loss-Hupeng}.

\subsection{Loss functions for comparing images}
In tasks related to image comparing, various network architectures have been extensively explored while loss functions remain relatively under-developed. 
There are primarily two categories of loss functions commonly used: (1) pixel-wise loss functions such as L1~\cite{l1_loss} and L2 loss~\cite{L2_loss}. 
These types of loss functions operate at the level of individual pixels or elements in the output and target images.
(2) global loss functions such as Gram loss~\cite{Gram_loss}, adversarial loss (GAN)~\cite{GAN_loss}, perceptual loss\cite{zhang2018LPIPS}, and SSIM loss~\cite{ssim, s3im}. 
In contrast to pixel-wise loss functions, global loss functions consider the relationships and structures across the entire image.
There has been research~\cite{loss_fucntion_review} that indicates the quality of the results improves significantly with better loss functions, empirically a combination of pixel-wise loss and global loss. 
However, there is currently no effective loss function that can capture information about non-semantic image context. 

\section{Method}
In this section, we formulate a novel non-semantic image descriptor of image context, \textbf{Glance Vector}, and a novel Glance Similarity Index Measure, \textbf{GlanceIM}, that benefits the training image reconstruction network.
\subsection{Glance Vector} 
Given an image \( \mathbf{I} \in \mathbb{R}^{h \times w} \), we mimic the human recognition process that bypasses the semantic concept with a set of global and local Glance Vectors.
\subsubsection{Global Glance Vector}
Firstly, we retrieve the global image context from $I$. A set of \( n \cdot m \) pixels, where \( n\cdot m < h\cdot w \) , is randomly selected from \( \mathbf{I} \), denoted by \( S \):
\[
S = \{\mathbf{I}_{ij} \mid (i, j) \in \Omega\}
\]
where \( \Omega \subseteq \{1, \dots, h\} \times \{1, \dots, w\} \) is the set of coordinates corresponding to the selected pixels. Secondly, a Glance vector is formulated by randomly retrieving \( n_g \cdot m_g \) pixels from \( S \) and forming a vector of shape \( \mathbf{v}_l \in \mathbb{R}^{n_g} \). 

We extract Glance vectors in a window-based computation: reshape $S$ into a 2D matrix, $\mathbf{S}$, of shape $n \times m$ and apply a 2D window of shape $n_g \times m_g$, resulting in $L_G$ submatrices, $\{\mathbf{V}_l|l = 1,..., L_G\}$. Different from SSIM\cite{ssim} and S3IM \cite{s3im}, the kernel is uniform, instead of circular-symmetric Gaussian weighting, since the stochastic global context is a group-based term and should not have a focus center. Moreover, we apply a unit stride to form a dense representation of the image context. The corresponding Glance vectors are obtained by flattening each submatrix $\mathbf{V}_l$ into a one-dimensional vector $\mathbf{v}_l$. The dense set of global Glance Vectors is represented as:
\[
\mathcal{V}_{Global} = \{\mathbf{v}_l \in R^{n_g m_g} \mid l = 1, \dots, L_G\}
\]
\( \mathcal{V}_{Global} \) provides a more compact and computationally efficient representation of the global context embedded in \( \mathcal{S} \).

The construction of \( \mathcal{V}_{Global} \) can leverage prior knowledge of human perception by translating it into a pixel selection rule that emphasizes perceptually important structure. For instance, humans perceive bright and high-contrast areas more effectively than dim and low-contrast ones\cite{perception_contrast}. This also extends to well-trained physicians in clinical settings, where imaging modalities are physically designed to capture critical regions with high contrast, as higher-intensity areas often correlate with diagnostically significant regions. Below, we present an example of integrating Glance with perceptual priors to enhance undersampled MRI reconstruction.

\noindent\textbf{MRI Air Prior.}
The air region appears as a black area in the MRI image and the physicians don't consider them for diagnosis. This is because MRI works by detecting the magnetic signals produced by hydrogen nuclei in water molecules. Since air typically does not contain water molecules, the MRI scanner can not detect any meaningful magnetic signal produced by hydrogen nuclei in the air. We can simulate this perception process by not sampling $\mathcal{S}$ from air pixels.

\begin{figure}[t]
    \centering
    \includegraphics[width=0.95\linewidth]{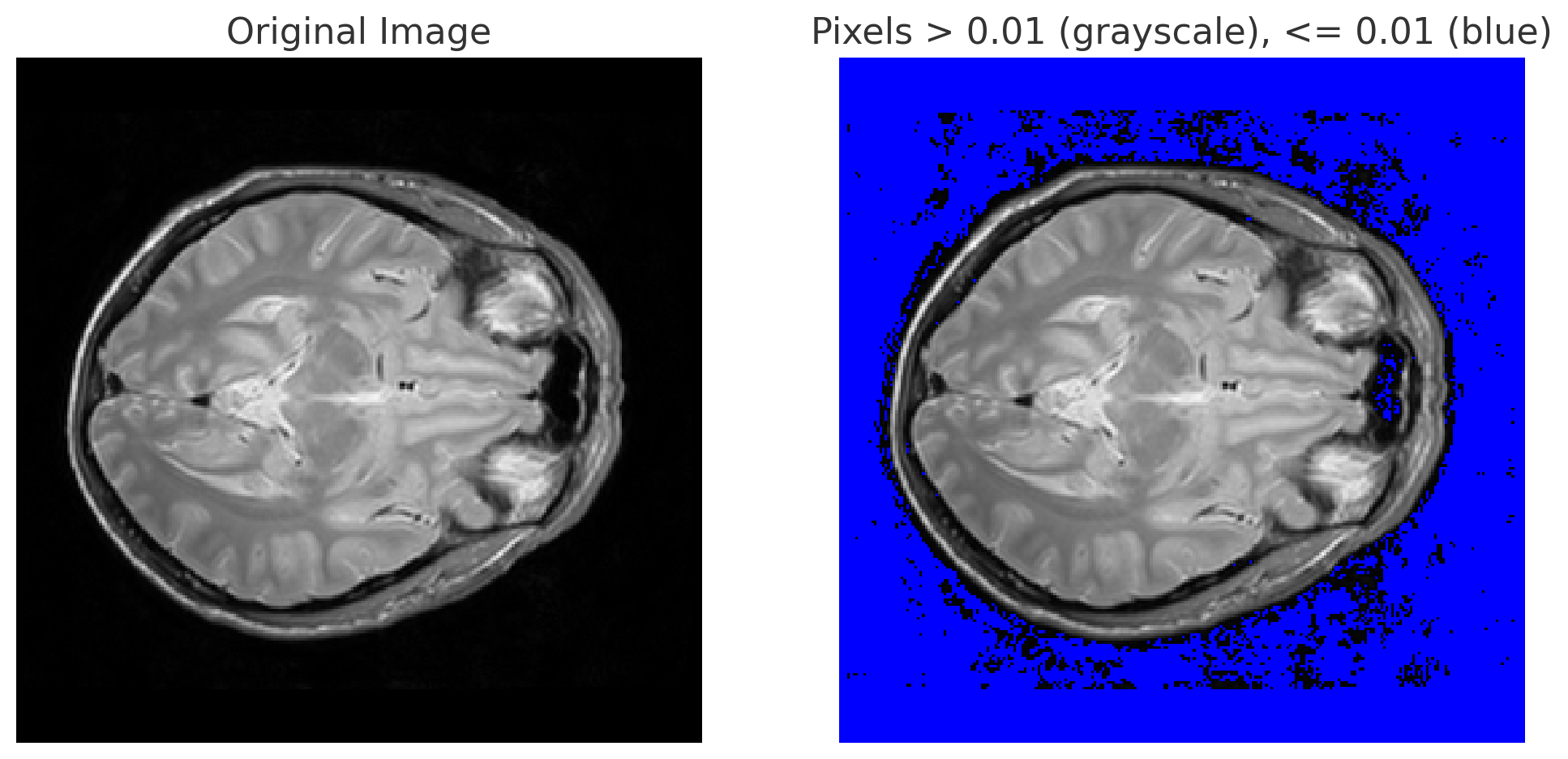}
    \caption{Leverage the MRI air prior with an intensity threshold.}
    \label{fig:air}
\end{figure}

As shown in Fig. \ref{fig:air}, we can define a small intensity threshold $\delta$, say 0.01, and sample the pixels whose intensity is higher. Formally, this process can be defined as
\[
S = \{ \mathbf{I}_{ij} \mid (i, j) \in \Omega \cap \mathbf{I}_{ij} > \delta \}.
\]

\subsubsection{Local Glance Vector}
Dedicated image reconstruction requires a closer look at the local image context. As such, we additionally extract a set of local Glance vectors from the original image $\mathbf{I}$. We apply the 2D uniform window of shape $n_g \times m_g$ to $I$ and produce a set of local Glance Vector by flattening. The dense set of local Glance Vectors is represented as:
\[
\mathcal{V}_{Local} = \{\mathbf{v}_l \in R^{n_g m_g} \mid l = 1, \dots, L_{L}\}
\]

Multi-scale Glance (MS-Glance) Vectors $\mathcal{V}$ are defined as the union of Global and Local Glance Vectors:
\[
\mathcal{V} = \mathcal{V}_{Local} \cup \mathcal{V}_{Global}.
\]
\subsection{Glance Index}

Given two Glance vectors \( v_l \) and \( v_{l'} \) from $\mathcal{V}$, the similarity between them is defined as the dot product of their standardized versions. The standardization of each vector involves subtracting the mean and dividing by the standard deviation of its elements. Specifically, the normalized Glance Vectors are denoted as \( \tilde{v}_l \) and \( \tilde{v}_{l'} \), where
\[
\tilde{\mathbf{v}}_l = \frac{\mathbf{v}_l - \mu_{\mathbf{v}_l}}{\sigma_{\mathbf{v}_l}}, \quad \tilde{\mathbf{v}}_{l'} = \frac{\mathbf{v}_{l'} - \mu_{\mathbf{v}_{l'}}}{\sigma_{\mathbf{v}_{l'}}}.
\]
Here, \( \mu_{\mathbf{v}_l} \) and \( \sigma_{\mathbf{v}_l} \) represent the mean and standard deviation of the elements in vector \( \mathbf{v}_l \), respectively. The similarity \( S(\mathbf{v}_l, \mathbf{v}_{l'}) \) between them is given by:
\[
S(\mathbf{v}_l, \mathbf{v}_{l'}) = \tilde{\mathbf{v}}_l \cdot \tilde{\mathbf{v}}_{l'}
\]
This Glance similarity measure reflects the correlation between the two vectors after accounting for their respective means and variances.

The normalized Glance Vectors \( \tilde{\mathbf{v}}_l \) and \( \tilde{\mathbf{v}}_{l'} \) take values in \( \mathbb{R}^{n_v \times m_v} \), where each element \( \tilde{v}_{l,i} \) follows a standard normal distribution \( N(0, 1) \). The inner product \( S(\tilde{\mathbf{v}}_l, \tilde{\mathbf{v}}_{l'}) \) of the normalized Glance Vectors lies within the range:
\[
-1 \leq S(\tilde{\mathbf{v}}_l, \tilde{\mathbf{v}}_{l'}) \leq 1
\]
The maximum value is achieved when the vectors are perfectly aligned, and the minimum value is achieved when they are perfectly anti-aligned.

From an algebraic perspective, The similarity of $(\mathbf{v}_l$ and $\mathbf{v}_{l'})$ is equal to a normalized covariance term:
From an algebraic perspective, the similarity of \( \mathbf{v}_l \) and \( \mathbf{v}_{l'} \) is equal to a Pearson correlation coefficient:
\begin{align}
    S(\mathbf{v}_l, \mathbf{v}_{l'}) &= \frac{\mathbf{v}_l - \mu_{\mathbf{v}_l}}{\sigma_{\mathbf{v}_l}} \cdot \frac{\mathbf{v}_{l'} - \mu_{\mathbf{v}_{l'}}}{\sigma_{\mathbf{v}_{l'}}} \\
    &= \frac{1}{\sigma_{\mathbf{v}_l} \cdot \sigma_{\mathbf{v}_{l'}}} \sum_{i=1}^{n_v \cdot m_v} \left[(v_{l,i} - \mu_{\mathbf{v}_l})(v_{l',i} - \mu_{\mathbf{v}_{l'}})\right] \\
    &= \frac{\text{Cov}(\mathbf{v}_l, \mathbf{v}_{l'})}{\sigma_{\mathbf{v}_l} \cdot \sigma_{\mathbf{v}_{l'}}}
\end{align}

We add a small constant \( C_s \) to both the numerator and denominator\cite{ssim} of the Glance Index to avoid numerical instability.

\subsection{Glance Index Measure}
Given two images \( I_0 \) and \( I_1 \), their Glance Index Measure (GlanceIM) is defined as the average of the Glance Index over two sets of MS-Glance Vectors, \(\mathcal{V}_0 \) and \( \mathcal{V}_1 \), which are extracted from two images in the same way:
\[
\textbf{GlanceIM}(I_0, I_1) = \frac{1}{|\mathcal{V}_0|} \sum_{v_0 \in \mathcal{V}_0, v_1 \in \mathcal{V}_1} \mathbf{S}(v_0, v_1)
\]
where \( \mathcal{V}_0 \) and \( \mathcal{V}_1 \) are the sets of Glance Vectors randomly sampled in the same manner from images \( I_0 \) and \( I_1 \), respectively.
\( \textbf{GlanceIM}(I_0, I_1) \) lies within the range \((-1, 1)\), where a value of 1 indicates perfect similarity (the images are identical in terms of global structure), and -1 indicates complete dissimilarity (the images are maximally different).

GlanceIM can be used as a loss for supervising image restoration networks by changing its range into $[0,2]$:
\[
L_{Glance}(I_0, I_1) = 1 - \textbf{GlanceIM}(I_0, I_1). 
\]

\section{Applications}
This section demonstrates how MS-Glance improves supervised image reconstruction. We first choose a simple regression task, fitting an image with INR then demonstrate the applicability of MS-Glance and the novel air prior in undersampled MRI reconstruction, spanning various acquisition scenarios and organs.

\noindent\textbf{Glance Loss Implementation.}
We validate MS-Glance and decompose it MS-Glance into Local Glance and Global Glance, which represent the Glance Vector sets used for computing GlanceIM. For 3-channel RGB images, the Glance vectors are defined as flattened 1D vectors of shape $3 \cdot n_g \cdot m_g$ to leverage the correlation information among channels. For 2-channel complex-valued MRIs, the Glance vectors are extracted from the image magnitude, which is the root-sum-of-square of the two channels that represent the real and imaginary parts. This operation allows us to directly incorporate MRI air prior when constructing $V_{Global}$. We set $n_g=m_g=16$, $n=m=96$, and $C_s=0.03$.

\noindent\textbf{Comparsion and Evaluation.} For comparison with existing losses, we add several common losses for training image restoration networks, including a feature-based loss, Perceptual Loss (LPIPS)\cite{zhang2018LPIPS}, a local-structure-emphasized loss, SSIM Loss\cite{ssim}, and Stochastic SSIM (S3IM) \cite{s3im}, a recent loss used in NeRF for multi-view synthesis. We use the official implementation of LPIPS\footnote{https://github.com/richzhang/PerceptualSimilarity} and S3IM\footnote{https://github.com/Madaoer/S3IM-Neural-Fields} and follow their default settings. For SSIM, we keep the default setting in Pytorch\footnote{https://github.com/Po-Hsun-Su/pytorch-ssim} and set its local window size to 16, the same as our $n_g$ and $m_g$, and stride to 1. This gives us a direct comparison of our Local Glance and SSIM - two metrics that both focus on local regions. The image reconstruction performance is evaluated with peak signal-to-noise ratio (PSNR) and structural similarity index measure (SSIM). 

All code is written in PyTorch and experiments are carried on an NVIDIA 3090 GPU with 24 GB memory. During training, gradient clipping is deployed for all.

\subsection{Image Fitting with INR}

\begin{table*}[t]
    \centering
\begin{tabular}{l|cc|cc|cc}
   \toprule
   Dataset & \multicolumn{2}{c|}{Coco - Objects} & \multicolumn{2}{c|}{CelebA - Human Face} & \multicolumn{2}{c}{IXI - Structural MRI }  \\
   \midrule 
    & PSNR(dB)   & SSIM  & PSNR(dB) & SSIM & PSNR(dB)  & SSIM  \\
   SIREN\cite{sitzmann2019siren} & 34.469 & 0.9396 & 39.007 & 0.9696 & 37.430 & 0.9743 \\
   +LPIPS\cite{zhang2018LPIPS} & 34.803 & 0.9417 & 39.346 & 0.9714 & 33.506 & 0.9288  \\
   +SSIM\cite{ssim} & 34.939 & 0.9468 & 39.232 & 0.9711 & 31.370 & 0.9298 \\
   +S3IM\cite{s3im} & 34.717 & 0.9430 & 39.293 & 0.9714 & 37.412 & 0.9604 \\
   +Local Glance & 35.004 & 0.9463 & 39.422 & 0.9723 & 39.088 & 0.9788 \\
   +Global Glance & 34.843 & 0.9439 & 39.329 & 0.9712 & 37.613 & 0.9601 \\
   +MS-Glance & \textbf{35.249} & \textbf{0.9493} & \textbf{39.571} & \textbf{0.9733} & \textbf{39.141} & \textbf{0.9788} \\
   \bottomrule
\end{tabular}
    \caption{Quantitative results of SIREN reconstruction with various loss functions on Coco, CelebA, and IXI datasets.}
    \label{tab:siren}
\end{table*}

\begin{figure}[t]
    \centering
    \includegraphics[width=0.95\linewidth]{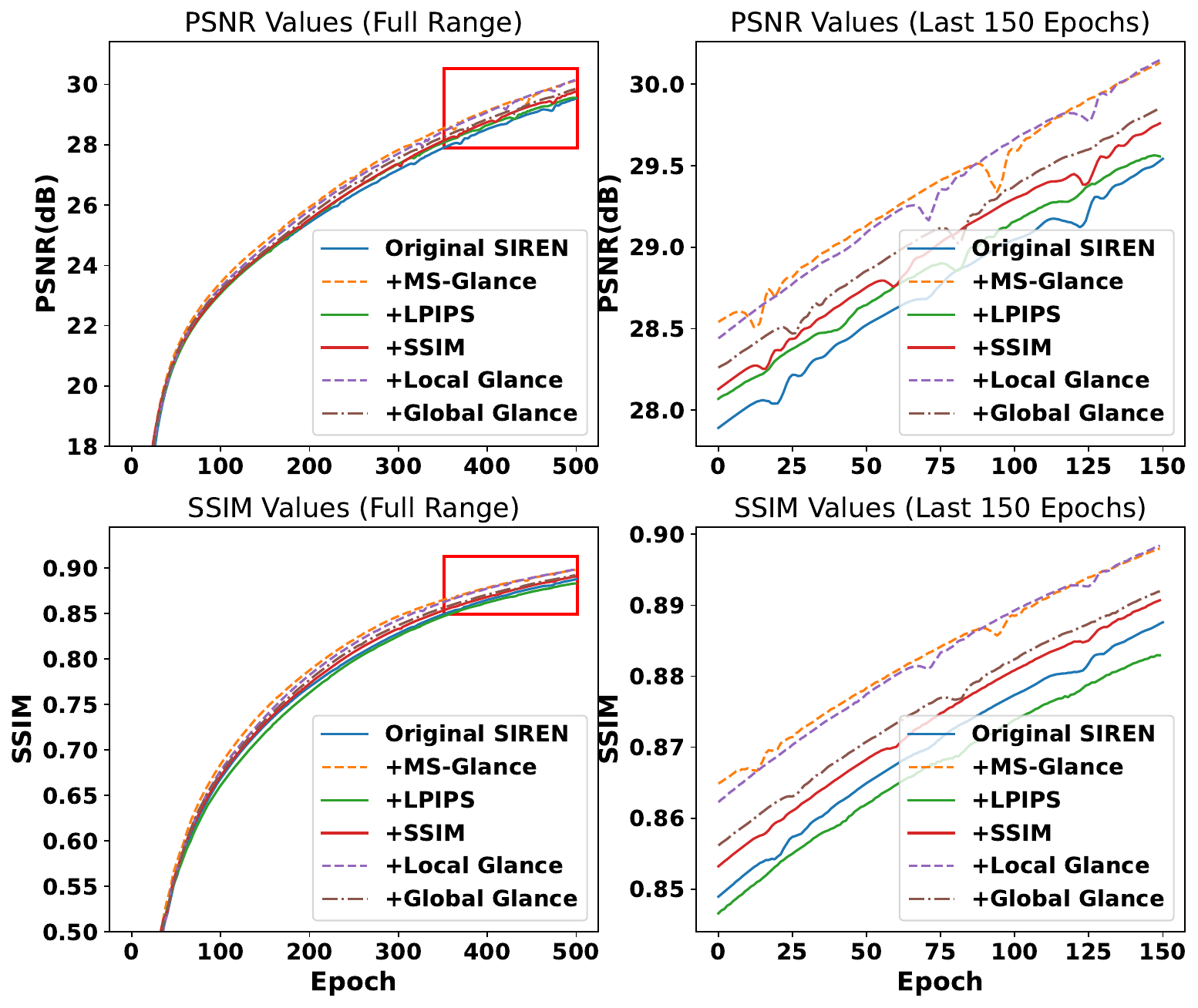}
    \caption{Step-wise SIREN reconstruction performance with various loss functions on a classic RGB image. The images on the right provide a zoomed-in view of the last 150 steps.}
    \label{fig:time}
    \vspace{-10pt}
\end{figure}

\begin{figure*}[t]
    \centering
    \includegraphics[width=0.9\linewidth]{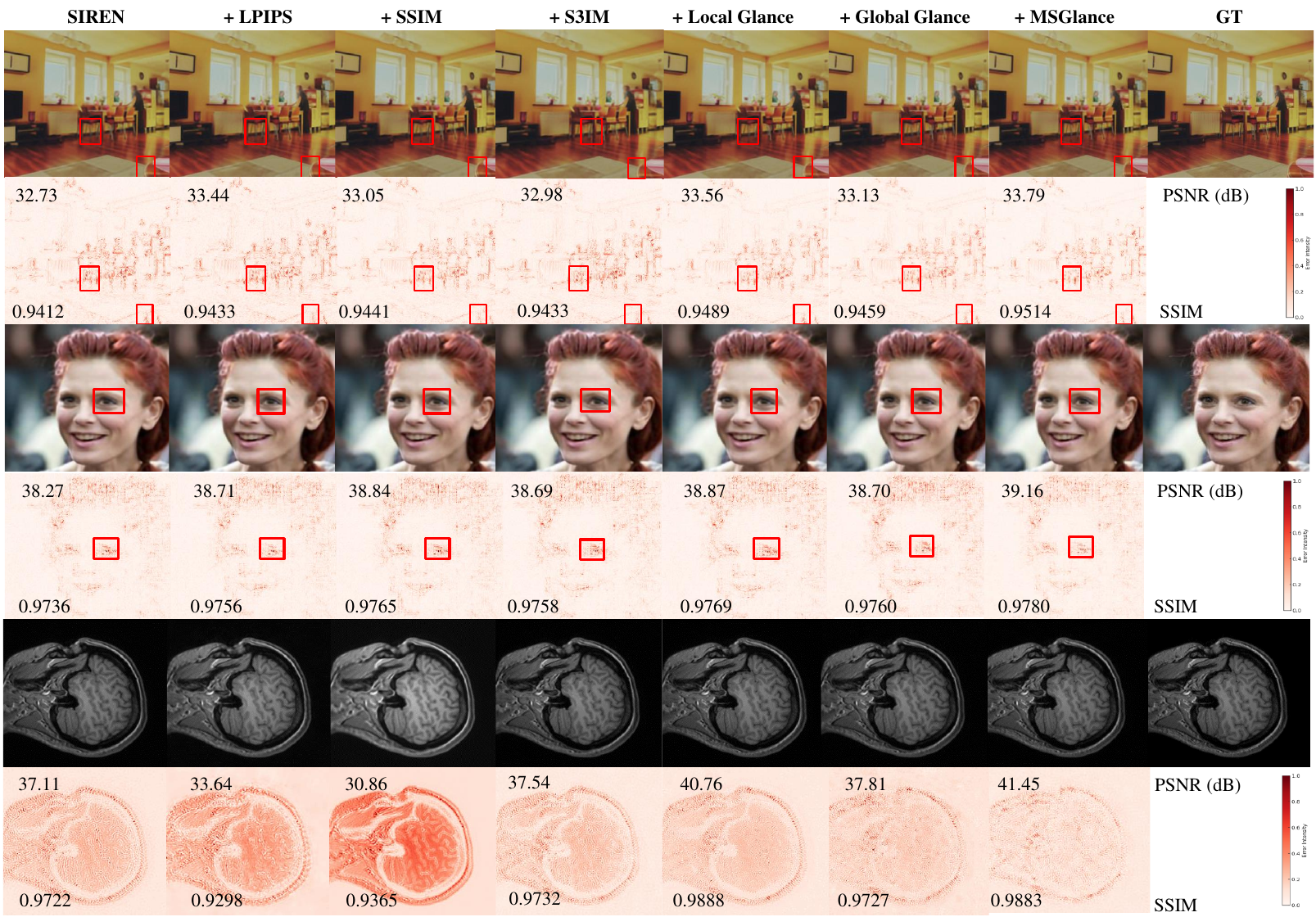}
    \caption{Qualitative results of SIREN reconstruction with various loss functions on Coco, CelebA, and IXI datasets. The odd rows show the reconstructed images and the even rows show the corresponding error maps, which are computed as the mean absolute value between the reconstructed image and the ground truth (GT). Error maps are normalized per row for better visualization. }
    \label{fig:SirenFitting}
    \vspace{-5pt}
\end{figure*}

In this section, we show how MS-Glance improves network reconstruction ability over existing losses based on an implicit neural network, SIREN\cite{sitzmann2019siren}.

\noindent\textbf{Model Implementation.}  SIREN's input is a Fourier-encoded 2D coordinate, and the output is the pixel intensity: 3 channels for RGB images and 1 channel for gray-scale images. Its architecture is a Multi-Layer Perceptron (MLP) network with 256  hidden features. It is originally trained with an L2 loss. We use the official implementation\footnote{https://github.com/vsitzmann/siren/blob/master/explore\_siren.ipynb} of SIREN and follow its default configuration for training: Adam optimizer combined with a learning rate of 0.0001.

\noindent\textbf{Dataset.} We use multiple categories of images from three public datasets: COCO\cite{lin2015microsoftcoco}, CelebA\cite{liu2015celebA} and IXI\footnote{https://brain-development.org/ixi-dataset/}. They contain common objects in context, human faces in the wild, and T1-w MRI brains, respectively. Images are center-cropped into the shape of $224\times224$ and normalized to [0,1]. We randomly choose 100, 100, and 1000 images from the three datasets, fit each image three times, and evaluate the PSNR and SSIM of the reconstructed image. 

\noindent\textbf{Loss implementation.} All compared losses are scaled with a coefficient of 0.01 to match the order of magnitudes of the L2 losses before combining with L2. For our MS-Glance and Global Glance, we don't assume any prior when selecting $\mathcal{V}_{Global}$. 

\noindent\textbf{Result.} The qualitative result is shown in Fig. \ref{fig:SirenFitting}.  LPIPS enhances natural image reconstruction but fails in the case of MRI. Both SSIM and Local Glance improve local structures, highlighted by the red boxes around the room and the woman's double-fold eyelids. However, SSIM introduces color distortion in the MRI reconstruction. S3IM and Global Glance improve overall image reconstruction, with Global Glance showing fewer losses, particularly evident in the error map of the last row. MSGlance combines the strengths of Local and Global Glance, achieving better structural reconstruction. The quantitative performance is shown in Table \ref{tab:siren}, further demonstrating the effectiveness of MS-Glance over others in all three scenarios. Specifically, local and global Glance show improvement separately and compensate for each other in MS-Glance.

We further visualize the SIREN regression process supervised by MS-Glance and other methods. Using a classic RGB image of shape $512 \times 512$, \textit{Astronaut}\footnote{https://scikit-image.org/docs/stable/api/skimage.data.html}, we plot PSNR and SSIM at each step of the SIREN regression, as shown in Fig. \ref{fig:time}. MS-Glance consistently outperforms existing loss functions throughout the entire training process. Additional qualitative results using \textit{Astronaut} are provided in the supplementary materials.
\subsection{Undersampled MRI reconstruction}
In this section, we show the effectiveness of Glance in a real-world application, undersampled MRI reconstruction. We use a popular recurrent learning backbone, Dilated Residual Dense Network (DRDN)\cite{frequency_loss2} for training and use the standard loss for training image-domain MRI reconstruction networks, L1\cite{zbontar2018fastMRI}, as the baseline.
\label{app:MRI}
\begin{table*}[]
    \centering    \resizebox{0.8\textwidth}{!}{%
\begin{tabular}{l|cc|cc|cc|cc}
            \toprule
            Dataset     & \multicolumn{4}{c|}{IXI - Brain}                         & \multicolumn{4}{c}{FastMRI - Knee}                     \\
                 \midrule
             Acceleration Rate    & \multicolumn{2}{c|}{5x} & \multicolumn{2}{c|}{7x} & \multicolumn{2}{c|}{5x} & \multicolumn{2}{c}{7x} \\
                 \midrule
                 & PSNR (dB)       & SSIM      & PSNR (dB)      & SSIM      & PSNR (dB)      & SSIM      & PSNR (dB)     & SSIM      \\
    L1               & 30.973     & 0.9512    & 30.651     & 0.9481    & 33.196     & 0.9068    & 31.117     & 0.8762    \\
    +LPIPS\cite{zhang2018LPIPS}         &   30.831         &    0.9496       &   30.262       &    0.9440       & 32.671     & 0.8974    & 30.704    & 0.8577    \\
    +SSIM\cite{ssim}          & 30.704     & 0.8577    & 30.592     & 0.9485    & 33.191     & \textbf{0.9078}    &     31.363       &  0.8835         \\
    +S3IM\cite{s3im}          & 31.017     & 0.9513    & 30.681     & 0.9484    & 33.236    &  0.9071    &    31.344  &  0.8794   \\
    +Global Glance & 31.122     & 0.9524    & 30.711     & 0.9485    &    33.253
&    0.9073     &      31.454      &    0.8804       \\
    +Local Glance  & 31.346     & 0.9537    & 30.813     & 0.9483    & 33.187        &  0.9038         &   31.352 &  0.8776  \\
    +MS-Glance     &    \textbf{31.434}     &   \textbf{0.9535}  &   \textbf{30.865}    &  \textbf{0.9485}   &   \textbf{33.299}        &   0.9072      &     \textbf{31.476}     &     \textbf{0.8783}      \\     
    \bottomrule
    \end{tabular}}
    \caption{Quntitative results of undersampled MRI reconstruction with various loss functions on Coco, CelebA, and IXI datasets.}
    \label{tab:mri_matrix}
    \vspace{-5pt}
\end{table*}

\noindent\textbf{Model Implementation.}
DRDN's input is a 2-channel undersampled MRI image and the output is a 2-channel reconstructed image. It stacks densely connected atrous layers and thus has a large receptive field while preserving image details. We use the official implementation\footnote{https://github.com/bbbbbbzhou/DuDoRNet} and change its recurrent number into 3 due to the GPU memory limit. DRDN is trained with an Adam optimizer and a learning rate of 0.0001 for 40 epochs for IXI and 20 epochs for FastMRI, and the model checkpoints used for evaluation are the one that produces the best PSNR on the validation set.

\noindent\textbf{Dataset and simulation.}
We use Proton Density weighted (PD-w) MRI from one simulation dataset, the IXI brain dataset, and one raw dataset, the FastMRI\cite{zbontar2018fastMRI} knee dataset. 
For IXI, we take 576 Volumes and uniformly sample slice-wise and then split volume-wise, resulting in 2455 slices for training and 700 slices for testing (matrix size = 256$\times$256). The FastMRI dataset includes 1,172 volumes of single-coil complex-valued PD-w k-space. We use the official dataset split and drop the first and last five noisy slices from each volume, resulting in 25012 training slices and 5145 testing slices (matrix size = 320$\times$320). Image normalization is done by normalizing image magnitude into [0,1].

\begin{figure*}[t]
    \centering
    \includegraphics[width=0.9\linewidth]{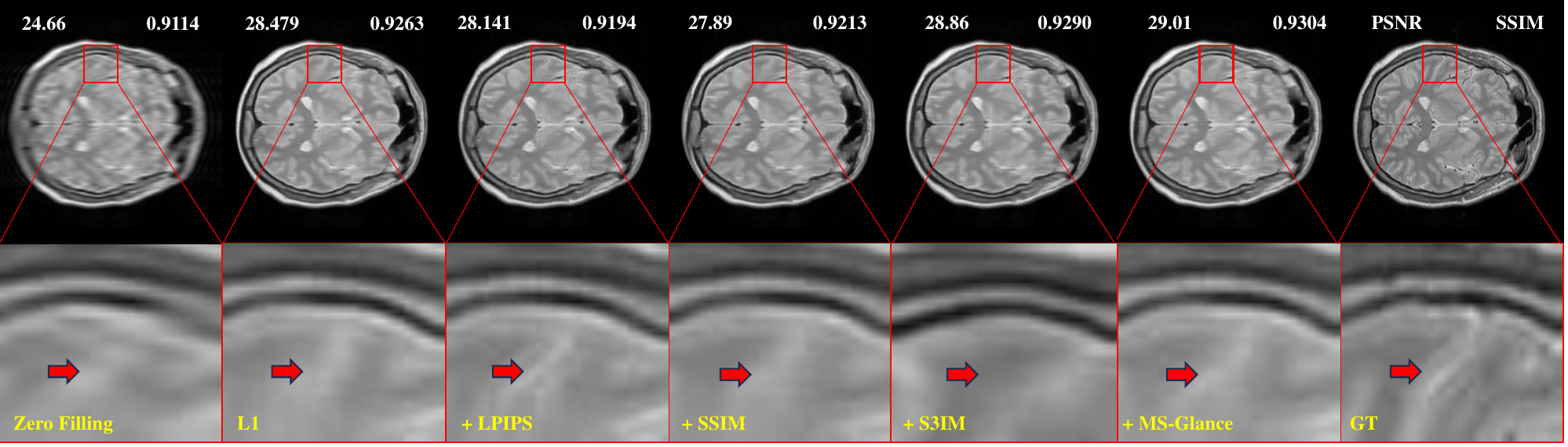}
    \caption{Conventional loss functions for training undersampled MRI reconstruction networks compared with our MS-Glance. The reconstructed images are evaluated with PSNR and SSIM and are zoomed in at the red region, which contains the cortex structure.}
    \label{fig:mrirecon}
    \vspace{-10pt}
\end{figure*}

We simulate the Uniform 1D undersampling using the random mask generation function from the official implementation of \cite{zbontar2018fastMRI}. The undersampled images are obtained by applying the Fourier Transform to the corresponding ground truth images, masking with a randomly generated uniform 1D mask, and applying the inverse Fourier Transform. We train DRDN under multiple acceleration rates, including $5\times$ and $7\times$, and evaluate under the corresponding rates. The Auto-Calibration region ratio is set to $12.5\%$. We show a 7x undersampled image (Zero Filling) in Fig. \ref{fig:mrirecon}. It is reconstructed directly from undersampled k-space by inverse Fourier Transform without any learning. It contains strong visual aliasing artifacts and is blurred.

\noindent\textbf{Loss implementation.}
 All compared losses are scaled with a coefficient of 0.1 to match the order of magnitudes of the L2 losses before combined with L2. For our MS-Glance and Global Glance, we use air prior when selecting $\mathcal{V}_{Global}$ and set $\sigma$ as 0.01. 
 
\noindent\textbf{Result.} Fig. \ref{fig:mrirecon} presents the qualitative results. The red box highlights a portion of the brain cortex, indicated by the red arrow. Among all, only MS-Glance and LPIPS successfully reconstruct the structure from the blurry input. However, LPIPS tends to introduce hallucinations, such as the vertical line on the right of the `+LPIPS' text which is not in the Ground Truth. The quantitative results are provided in Table \ref{tab:mri_matrix} and show similar behaviors to INR image fitting. MS-Glance achieves the best PSNR among all.

\section{Ablation studies}
\begin{table}[]
    \centering
    \resizebox{0.8\columnwidth}{!}{%
    \begin{tabular}{c|ccc}
    \toprule
        $n\cdot m$&  PSNR (dB) & SSIM & Time (s) \\
        \midrule
        $32^2$ & 34.744 & 0.9430 & 13.37\\
        $96^2$ (ours) & \textbf{35.249} & \textbf{0.9493} & 20.71\\
        $128^2$ & 35.119 & 0.9470 & 29.91\\
        $224^2$ (whole) & 35.029 & 0.9464 & 76.78\\
    \bottomrule
    \end{tabular}}
    \caption{Ablation study of $n \cdot m$.}
    \label{tab:nm}
    \vspace{-10pt}
\end{table}

\begin{table}[]
    \centering
    \resizebox{0.8\columnwidth}{!}{%
    \begin{tabular}{c|ccc}
    \toprule
        $n_g\cdot m_g$&  PSNR (dB) & SSIM & Time (s) \\
        \midrule
        $4^2$ & 35.096 & 0.9474 & 9.94 \\
        $8^2$ & 35.247 & 0.9491 & 10.19\\
        $16^2$ (ours) & \textbf{35.249} & \textbf{0.9493} & 20.71\\
        $32^2$ & 35.188 & 0.9485 & 54.77\\
    \bottomrule
    \end{tabular}}
    \caption{Ablation study of $n_g \cdot m_g$.}
    \label{tab:nm_g}
    \vspace{-10pt}
\end{table}

\begin{table}[]
    \centering
    \resizebox{0.9\columnwidth}{!}{%
    \begin{tabular}{l|cc|cc}
    \toprule
       Acc. rates & \multicolumn{2}{c|}{5x} & \multicolumn{2}{c}{7x} \\
    \midrule
         & PSNR (dB) & SSIM & PSNR (dB) & SSIM \\
        Ours & 31.434 & 0.9535 & 30.865 & 0.9485\\
        w/o air prior & 31.308 & 0.9519 & 30.693 & 0.9478\\
    \bottomrule
    \end{tabular}
    }%
    \caption{Ablation study of the MRI air prior.}
    \label{tab:air_prior}
    \vspace{-10pt}
\end{table}

\subsection{The selection of $n \cdot m$ and  $n_g \cdot m_g$}
MS-Glance extracts the global context of an image by randomly retrieving a subset of $n \cdot m$ pixels. $n_g \cdot m_g$ is the shape of Glance vectors. We show the effect of using different values when fitting the INR on the CoCo dataset in Table \ref{tab:nm} and Table \ref{tab:nm_g} for $n \cdot m$ and  $n_g \cdot m_g$, respectively. Table \ref{tab:nm} shows that using the whole set of pixels is slow and does not lead to performance gain and our $n \cdot m$ selection leads to the best performance. Table \ref{tab:nm_g} shows that our selection of $n_g \cdot m_g$ achieves the best performance while $8^2$ achieves the best speed-performance trade-off.

\subsection{Effectiveness of the rule-based pixel selection}
The construction of $\mathcal{V}_{Global}$ can integrate perceptual prior, and we have shown an example of using MRI air prior. To validate the effectiveness of MRI air prior, we evaluate the performance of training undersampled MRI networks with and without it. The experiments are carried out on the FastMRI dataset under two acceleration rates. Table \ref{tab:air_prior} presents the performance of MS-Glance with and without the MRI air prior. This validates its beneficial effect.
\vspace{-5pt}

\section{Discussion}
\subsection{MS-Glance and SSIM}
SSIM~\cite{ssim} computes image similarity window-wise and emphasizes the center pixel with a Gaussian kernel. Given two windows $x$ and $y$, their similarity is the product of three statistics: luminance $l$, contrast $c$, and structure $s$:
\[
    l(x,y)={\frac {2\mu _{x}\mu _{y}+c_{1}}{\mu _{x}^{2}+\mu _{y}^{2}+c_{1}}},c(x,y)={\frac {2\sigma _{x}\sigma _{y}+c_{2}}{\sigma _{x}^{2}+\sigma _{y}^{2}+c_{2}}},
\]
\vspace{-5pt}
\[
    s(x,y)={\frac {\sigma _{xy}+c_{3}}{\sigma _{x}\sigma _{y}+c_{3}}},
\]
where $\mu _{x}, \mu _{y}$ are the mean of $x$ and $y$, $\sigma _{x}, \sigma _{y}$ are the variance of $x,y$ and $\sigma _{xy}$ is the covariance of $x,y$. $s$
 resembles the formulation of the Glance Index. However, the construction of Global Glance vectors gathers pixels across the image, rather than in local windows like SSIM. Additionally, when we combine the $l$ and $c$ of SSIM with our Glance Index, we witness a performance drop. We argue that the non-semantic image context we try to capture is highly correlated with image structure and thus is better measured by our proposed Glance Index unitarily. Moreover, MS-Glance does not assume locality or center for each window, since the image context is statistics-based and does not have a focus center. Ablation studies about the kernel selection can be found in the supplementary.
\subsection{Future Work}
MS-Glance incorporates multi-scale, non-semantic image context into supervised image reconstruction. It can be deployed in other low-level vision tasks such as super-resolution, deblurring, and compression. Also, the stochasticity property of Global Glance allows for integrating more perception priors. Future work includes combining MS-Glance with other explicit appearance priors\cite{Liu_2024_PEPSI,Liu_2023_BrainID} and learned implicit priors such as pixel correlation\cite{correlation,yan2024interpretable}.
\vspace{-5pt}
\section{Conclusion}
We have shown how to leverage the non-semantic image context, which can be captured by human vision, for supervising image reconstruction networks. We propose MS-Glance, a novel bio-inspired multi-scale descriptor of non-semantic image context and its loss form. We demonstrate its effectiveness in improving image reconstruction through two settings: INR fitting and undersampled MRI reconstruction. Finally, extensive experiments on both natural and medical datasets show that the MS-Glance loss outperforms existing loss functions used in image restoration. 
\vspace{-10pt}

\section{Acknowledgement}
This work is supported by Natural Science Foundation of China under Grant 62271465, Suzhou Basic Research Program under Grant SYG202338, and Open Fund Project of Guangdong Academy of Medical Sciences, China (No. YKY-KF202206).

{\small
\bibliographystyle{ieee_fullname}
\bibliography{egbib}
}

\end{document}


\title{Supplemenatry Materials of MS-Glance: Bio-Inspired Non-semantic Context Vectors and their Applications in Supervising Image Reconstruction}
\maketitle

Here, we show some additional results mentioned in the main paper: qualitative results on fitting the Astronaut image and ablation studies on the Glance's window kernel and distance measure. We also add additional details on MS-Glance's implementation and the network architecture of DRDN, which we use for undersampled MRI reconstruction experiments. 
\section{More qualitative results of \textit{Astronaut}}
\begin{figure*}[t]
    \centering
    \includegraphics[width=0.95\linewidth]{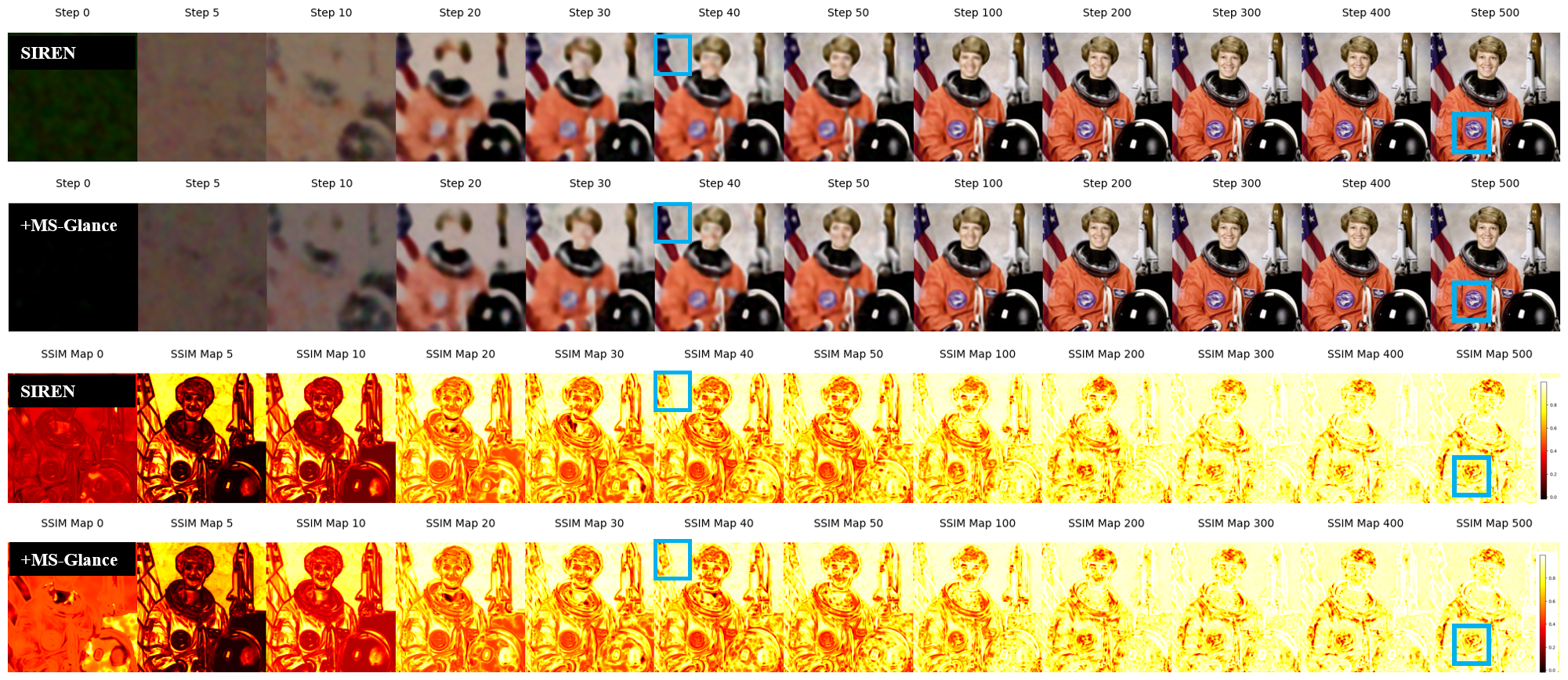}
    \caption{Step-wise reconstruction of the example image, \textit{Astronaut}}
    \label{fig:astro}
\end{figure*}

\textit{Astronaut} is a color image of the astronaut Eileen Collins. In Figure \ref{fig:astro}, we compare the step-wise reconstruction of \textit{Astronaut} by SIREN and SIREN+MSGlance. The reconstructed images and the corresponding SSIM error maps are visualized. MS-Glance reconstructs the image details faster (the blue boxes in step 40) and ends up with a finer reconstruction (the blue boxes in step 500).

\section{More ablation studies}
\subsection{Uniform Kernel and Gaussian Kernel}
The uniform window kernel is a key distinction between the Glance Index Measure and methods like SSIM and S3IM. To conduct a comprehensive ablation study, we replace our uniform kernel with their Gaussian kernel on both tasks. For MRI reconstruction, we use the IXI dataset under two acceleration rates. For INR fitting, we use the Coco dataset.

Table \ref{tab:kernel} states that compared with the Gaussian kernel, our uniform kernel not only stabilizes training but also enhances performance. Initially, when we applied the Gaussian kernel to MS-Glance, it caused the loss function to produce NaN values. To mitigate this, we detected NaN and switched to a standard $L_p$ loss during NaN iterations. However, a significant number of steps still resulted in NaN values. To further investigate, we decomposed MS-Glance. We observed that both Local Glance with Gaussian and MS-Glance with Gaussian led to approximately 60\% NaN loss, contributing to the large performance degradation. While the Global Glance with Gaussian's training remained stable, it also experienced a performance decline.

\begin{table}[h]
    \centering
    \resizebox{0.95\columnwidth}{!}{%
    \begin{tabular}{l|cc|cc|cc}
    \toprule
    & \multicolumn{4}{c|}{Undersampled MRI reconstruction} & \multicolumn{2}{c}{INR fitting}\\
       & \multicolumn{2}{c|}{5x} & \multicolumn{2}{c|}{7x} \\
    \midrule
         & PSNR (dB) & SSIM & PSNR (dB) & SSIM & PSNR (dB) & SSIM \\
    MS-Glance &  \textbf{31.434}     &   \textbf{0.9535}  &   \textbf{30.865}    &  \textbf{0.9485}   & \textbf{35.249} & \textbf{0.9493}\\
    MS-Glance - Gaussian & 29.497 & 0.9323 & 28.699 & 0.9229 & 35.099 & 0.9469\\
    \midrule
    Global Glance & \textbf{31.122}     & \textbf{0.9524}    & \textbf{30.711}     & \textbf{0.9485}  &  \textbf{34.843} & \textbf{0.9439} \\
    Global Glance - Gaussian & 31.104 & 0.9524 & 30.622 & 0.9483 & 34.827 & 0.9441 \\
    \midrule
    Local Glance  & \textbf{31.346}     & \textbf{0.9537}    & \textbf{30.813}     & \textbf{0.9483} & \textbf{35.004} & \textbf{0.9463}  \\
    Local Glance - Gaussian & 29.487 & 0.9315 & 28.573 & 0.9208 & 34.830 & 0.9430 \\
    \bottomrule
    \end{tabular}
    }%
    \caption{Ablation study of the window kernel.}
    \label{tab:kernel}
\end{table}

\subsection{Glance Index Measure and SSIM}
While we compare our method with SSIM loss in all experiments, we also highlight the connection between the Glance Index Measure and SSIM, which is discussed in detail in the main paper. In this section, we provide additional experimental results to compare the performance of the Glance Index Measure against SSIM. As mentioned in the main paper, the structural term of SSIM computes covariance similarly to how the Glance Index Measure operates. However, SSIM also incorporates luminance ($l$) and contrast ($c$) terms. To account for this, we extend our Glance Index Measure by integrating the computation of $l$ and $c$, multiplying them with the original Glance Index Measure. We tested this modified approach across both tasks. 

We perform the evaluation on both tasks. For MRI reconstruction, we use the IXI dataset under two acceleration rates. For INR fitting, we use the Coco dataset.
Table \ref{tab:LCS} demonstrates the effectiveness of the Glance Index Measure, particularly in global scenarios. The current Glance Index Measure shows that MS-Glance and Global Glance remain superior. However, the Local Glance enhanced with $l$ and $c$ exhibits improved performance, especially in SSIM computations. This improvement is expected, as it directly optimizes a term similar to SSIM itself. Additionally, we explored combining the original Global Glance design with the new Local Glance incorporating $l$ and $c$, with results shown in the last row. This approach, however, did not perform as well as the original MS-Glance design, suggesting a conflict between the two approaches.

\begin{table}[h]
    \centering
    \resizebox{0.95\columnwidth}{!}{
    \begin{tabular}{l|cc|cc|cc}
    \toprule
    & \multicolumn{4}{c|}{Undersampled MRI reconstruction} & \multicolumn{2}{c}{INR fitting}\\
       & \multicolumn{2}{c|}{5x} & \multicolumn{2}{c|}{7x} \\
    \midrule
         & PSNR (dB) & SSIM & PSNR (dB) & SSIM & PSNR (dB) & SSIM \\
    MS-Glance     &    \textbf{31.434}     &   \textbf{0.9535}  &  \textbf{30.865}    &  \textbf{0.9485} & \textbf{35.249} & \textbf{0.9493}\\
    + $l$ and $c$ of SSIM & 31.035 & 0.9523 & 30.660 & 0.9483 & 35.131 & 0.9484\\
    \midrule
    Global Glance (a) & \textbf{31.122}     & \textbf{0.9524}    & \textbf{30.711}     & \textbf{0.9485} & \textbf{34.843} & 0.9439\\
    + $l$ and $c$ of SSIM & 31.055 & 0.9523 & 30.653 & 0.9479 & 34.714 &\textbf{0.9426}\\
    \midrule
    Local Glance  & 31.346     & 0.9537    & 30.813     & 0.9483 & \textbf{35.004} & 0.9463  \\
    + $l$ and $c$ of SSIM (b) & \textbf{31.425} & \textbf{0.9557} & \textbf{30.939} & \textbf{0.9519} & 34.913 & \textbf{0.9468} \\
    \midrule
    Combination of (a) and (b) & 30.953 & 0.9516 & 30.701 & 0.9492 & 35.242 & 0.9496\\
    \bottomrule
    \end{tabular}
    }
    \caption{The effect of Glance Index Measure and SSIM to MS-Glance.}
    \label{tab:LCS}
\end{table}

\section{Additional Details}
\subsection{Implementation of MS-Glance}
In the Global Glance process, we randomly select pixels and shuffle them 10 times, resulting in more Glance vectors for computing the Global Glance Index Measure. As shown in Table \ref{tab:shuffleTime}, shuffling leads to a slight improvement in performance. The experiments are carried out on the Coco dataset.

\begin{table}[h]
    \centering
    \begin{tabular}{c|ccc}
        \toprule
        Shuffle times & 1 & 5 & 10\\
        \midrule
        PSNR & 35.202 & \textbf{35.257} & 35.249\\
        SSIM & 0.9489 & 0.9491 & \textbf{0.9493}\\
        \bottomrule
    \end{tabular}
    \caption{The effect of the shuffle time on INR fitting.}
    \label{tab:shuffleTime}
\end{table}

\subsection{Architecture of DRDN}
\begin{figure}[h]
    \centering
    \includegraphics[width=0.95\linewidth]{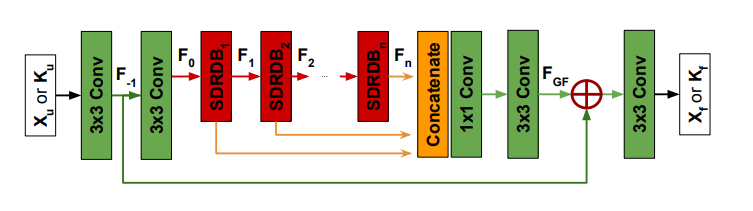}
    \caption{The diagram of DRDN\cite{frequency_loss2}.}
    \label{fig:drdn}
\end{figure}

We choose DRDN as the network for undersampled MRI reconstruction. Its strong performance has been validated by their original experiments and many recently established works\cite{chung2022scoreMRI-diff,song2021scorebased}. DRDN\cite{frequency_loss2} customizes the local and global structure design for the MRI reconstruction task. It uses a Squeeze-and-excitation Dilated Residual Dense Block (SDRDB) as the backbone. The main diagram is shown in Figure \ref{fig:drdn}.

Globally, DRDN consists of an initial feature extraction module (two sequential 3 × 3 convolution layers), multiple SDRDBs followed by global feature fusion (a concatenation operation for all SDRDBs' output), and global residual learning enhanced by a Squeeze-and-excitation on the residual branches.
\begin{figure}
    \centering
    \includegraphics[width=0.95\linewidth]{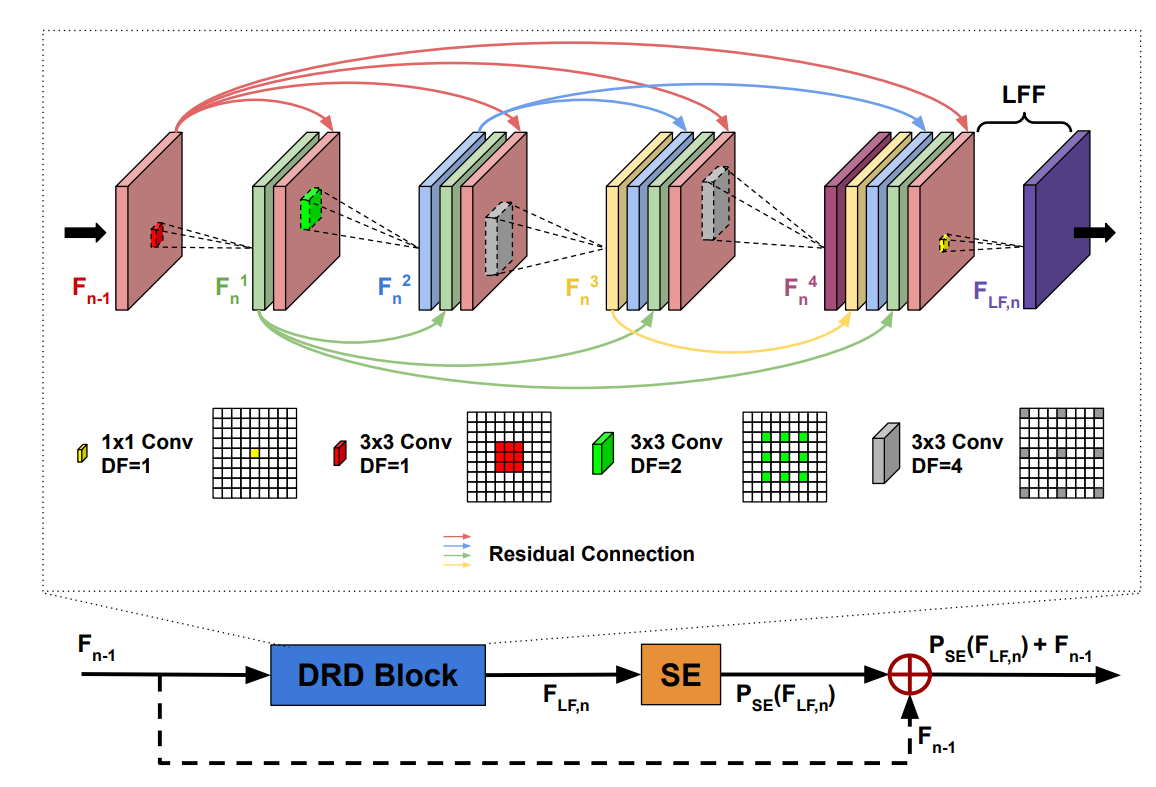}
    \caption{The component of each SDRDB.}
    \label{fig:sdrdb}
\end{figure}

The structure in each SDRDB is shown in Figure. In each SDRDB, there are four densely connected atrous convolution layers, local feature fusion, Squeeze-and-Excitation, and local residual learning. 

{\small
\bibliographystyle{ieee_fullname}
\bibliography{egbib}
}